\documentclass[english,aps,prd,nofootinbib,superscriptaddress,12pt]{revtex4-1}

\usepackage{graphicx} 

\usepackage[T1]{fontenc}
\usepackage{lmodern}
\usepackage{caption}
\usepackage{multirow}
\usepackage{floatrow}

\usepackage{subcaption}
\usepackage{hyperref}
\hypersetup{colorlinks,linkcolor={blue},citecolor={blue},urlcolor={blue}}
\usepackage[normalem]{ulem}

\usepackage{color}

\usepackage{mathtools}
\usepackage{physics}

\usepackage{amsmath}
\usepackage{amssymb}
\usepackage{slashed}
\usepackage{lineno}

\begin{document}

\title{Exploring the electromagnetic properties of neutrinos at a short-baseline reactor neutrino experiment}   

\date{\today}

\author{Guo-Fu~Cao}\email{caogf@ihep.ac.cn} \affiliation{Institute of High Energy Physics, Chinese Academy of Sciences, Beijing 100049, China} \affiliation{University of Chinese Academy of Sciences, Beijing 100049, China}
\author{Xin~Chen}\email{xchen@ihep.ac.cn} \affiliation{Institute of High Energy Physics, Chinese Academy of Sciences, Beijing 100049, China} \affiliation{School of Physics, Nanjing University, Nanjing 210093, China}
\author{Luis~A.~Delgadillo} \email{ldelgadillof@ihep.ac.cn}
\affiliation{Institute of High Energy Physics, Chinese Academy of Sciences, Beijing 100049, China}
\affiliation{Kaiping Neutrino Research Center, Guangdong 529386, China}  
\author{Maxim~Gonchar} \email{gonchar@jinr.ru} \affiliation{Joint Institute for Nuclear Research, Dubna, Russia} \affiliation{Institute for Nuclear Research of the Russian Academy of Sciences, Moscow, Russia}
\author{Yu-Feng~Li} \email{liyufeng@ihep.ac.cn}
\affiliation{Institute of High Energy Physics, Chinese Academy of Sciences, Beijing 100049, China} \affiliation{University of Chinese Academy of Sciences, Beijing 100049, China}
\author{Vitalii~Zavadskyi} \email{zavadskyi@jinr.ru} \affiliation{Joint Institute for Nuclear Research, Dubna, Russia} \affiliation{Institute for Nuclear Research of the Russian Academy of Sciences, Moscow, Russia}
\begin{abstract}
\noindent
Upcoming and present reactor neutrino experiments represent an appealing tool to probe fundamental properties in the neutrino sector. In this paper, we study the physics potential to determine the electromagnetic properties of neutrinos via electron--neutrino elastic scattering (E$\nu$ES) at a short-baseline neutrino experiment. We evaluate the sensitivity to the weak mixing angle, $\sin^2\theta_W$, employing antineutrinos from a nuclear reactor source. Furthermore, from the sensitivity to $\sin^2 \theta_W$, we obtain bounds on the neutrino charge radius. We also present the projected sensitivity to the effective neutrino magnetic moment, $\mu_{\nu}$. Compared with other reactor neutrino measurements, this experimental configuration may set competitive limits on the electromagnetic properties of neutrinos.
\end{abstract}

\maketitle

\section{Introduction}
\label{sec:intro}

The understanding of the three-neutrino oscillation paradigm has been made possible by establishing a nonzero reactor mixing angle ($\theta_{13}$) in reactor neutrino experiments, including Daya Bay~\cite{DayaBay:2012fng}, RENO~\cite{RENO:2012mkc}, and Double Chooz~\cite{DoubleChooz:2012gmf}. 
However, other interesting physical studies, including the electromagnetic properties of neutrinos, can be investigated in reactor neutrino experiments~\cite{Giunti:2008ve}. For instance, a proposal to determine the weak mixing angle ($\sin^2 \theta_{W}$) via neutrino-electron elastic scattering process at a reactor-based experimental configuration was assessed in Ref.~\cite{Conrad:2004gw}. An overview of the current status and future particle physics potential using reactor antineutrinos can be found elsewhere~\cite{Akindele:2024nzu}. 

In the Standard Model (SM), neutrinos are neutral under the electromagnetic (EM) interaction. However, a small charge radius can be generated via radiative corrections. For a thorough overview on the electromagnetic properties of neutrinos and current experimental status, we refer the reader to Refs.~\cite{Broggini:2012df, Giunti:2014ixa, Giunti:2024gec}. Bounds on the electromagnetic properties of neutrinos include: Krasnoyarsk~\cite{Vidyakin:1992nf}, Rovno~\cite{Derbin:1993wy}, LAMPF~\cite{Allen:1992qe}, LSND~\cite{LSND:2001akn}, TEXONO~\cite{TEXONO:2006xds, TEXONO:2009knm}, Bugey (MUNU)~\cite{MUNU:2005xnz}, Borexino~\cite{Borexino:2008dzn, Borexino:2017fbd}, GEMMA~\cite{Beda:2012zz}, CONUS~\cite{CONUS:2022qbb}, XENON1T~\cite{XENON:2020rca}, PandaX~\cite{PandaX-II:2020udv}, Dresden-II data-sets~\cite{Coloma:2022avw, AtzoriCorona:2022qrf}, COHERENT data-sets~\cite{Cadeddu:2020lky, DeRomeri:2022twg, AtzoriCorona:2024rtv}, as well as recent CE$\nu$NS measurements from CONUS+~\cite{Alpizar-Venegas:2025wor, Chattaraj:2025fvx, DeRomeri:2025csu, AtzoriCorona:2025ygn}. Besides, prospects to determine the electromagnetic properties of neutrinos have been considered in the context of long-baseline neutrino experiments such as DUNE~\cite{Miranda:2021mqb, Mathur:2021trm}. Recently, in Ref.~\cite{Herrera:2024ysj}, the authors explored the potential to probe the anapole moment of neutrinos at liquid xenon detectors.

In this paper, we focus on the elastic scattering of neutrinos with electrons; we obtain expected sensitivities to the weak mixing angle, neutrino charge radius, and magnetic moment of neutrinos considering the case of a short-baseline reactor anti-neutrino experiment~\cite{JUNO:2020ijm} located about $44$ m from a nuclear reactor core, with electron energy resolution of 2\% at 1 MeV. This setup is conceptually similar to the proposed TAO detector, which was discussed as part of the physics program for reactor neutrino studies~\cite{JUNO:2015sjr, JUNO:2015zny}. \footnote{For instance, initial proposals to construct a near detector at a short baseline for reactor neutrino experiments were discussed in Refs.~\cite{Li:2013zyd, Wang:2016vua, Forero:2017vrg, Capozzi:2020cxm}.}
In addition to providing a reference spectrum, other physics scenarios that can be probed at such a short-baseline facility include: light sterile neutrino searches~\cite{JUNO:2020ijm, Berryman:2021xsi, Basto-Gonzalez:2021aus}, light dark boson searches~\cite{Smirnov:2021wgi}, heavy neutral lepton searches~\cite{vanRemortel:2024wcf}, radiative corrections in the elastic neutrino-electron scattering channel~\cite{Brdar:2024lud}, as well as renormalization group effects on the Dirac $CP-$violating phase~\cite{Ge:2024ibn}.

The manuscript is organized as follows: In Section~\ref{sec:enues}, we present the formalism of electron--neutrino elastic scattering. In Section~\ref{sec:method}, we describe the experimental configuration of a representative short-baseline detector and the methodology. In Section~\ref{sec:results}, we show our results of the expected sensitivities to the weak mixing angle, neutrino charge radius, and neutrino magnetic moment at a 44-meter baseline setup. Finally, in Section~\ref{sec:conclusion}, we give our conclusions and comment on future assessments in this direction.

\section{Electron--neutrino elastic scattering}
\label{sec:enues}
Reactor neutrino experiments can probe the electromagnetic properties of neutrinos through the elastic scattering of antineutrinos with electrons ($\bar{\nu}_e-e^{-}$). Besides, this data-set can be employed to reduce the neutrino flux uncertainty~\cite{MINERvA:2022vmb}. Using this process, several reactor neutrino experiments have set competitive limits on the aforementioned properties of neutrinos, and these bounds are approaching the SM predictions~\cite{Vidyakin:1992nf, Derbin:1993wy, TEXONO:2006xds}. All of these experimental findings offer an exciting possibility to pursue searches that could uncover the electromagnetic properties of neutrinos in future reactor neutrino experiments.

The differential cross section for the electron antineutrino--electron elastic scattering process (E$\nu$ES) at tree-level in the SM reads~\cite{Vogel:1989iv}
\begin{equation}
\label{xsc}
    \frac{d\sigma(T_e, E_\nu)}{dT_e} = \frac{G_F^2 m_e}{2 \pi} \Big[ (g_V-g_A)^2 +(g_V+g_A)^2 \Big(1-\frac{T_e}{E_\nu} \Big)^2-\big(g_V^2-g_A^2 \big) \frac{m_eT_e}{E_\nu^2} \Big]~.
\end{equation}
Here, $T_e$ and $m_e$ are the recoil kinetic energy and mass of the electron, $G_F$ is the Fermi constant, $E_\nu$ is the neutrino energy, and $g_V= 2\sin^2 \theta_{W}+1/2,~g_A=-1/2$ are the corresponding vector ($V$) and axial ($A$) couplings, with $\theta_{W}$ the weak mixing angle. Regarding our sensitivity analysis, for the weak mixing angle, throughout this work as the reference value, we will consider the SM best-fit at low energy $\sin^2 \theta_{W} = 0.2385$~\cite{Tiesinga:2021myr}. However, aside from the contributions of the magnetic moment and neutrino charge radius, in this analysis, we will not consider other radiative corrections to the E$\nu$ES cross section~\cite{Brdar:2024lud, Tomalak:2019ibg, Huang:2024rfb}; their inclusion yields a negligible shift in the projected sensitivities within our systematic error budget (see Sec.~\ref{syst}).

\section{Experimental setup and methodology}
\label{sec:method}
We consider a representative short-baseline (SBL) detector located at a distance of around 44 m from a reactor core with a nominal thermal power of 4.6 GW. The experimental site is assumed to be on the ground floor of the reactor hall at a depth of 9.6 meters below ground level. In particular, the central detector consists of a 2.8-tonne spherical Gadolinium (Gd)-doped liquid scintillator (GdLS), and approximately 1 tonne of fiducial mass; it is equipped with an arrangement of silicon photo-multipliers (SiPMs)~\cite{JUNO:2020ijm}. This configuration provides a benchmark for studying the potential of short-baseline reactor experiments.

\subsection{Backgrounds}
\label{subsec:bckg}
The inverse beta decay (IBD) process dominates while the reactor is operating. However, this background can be monitored due to the tagging capabilities of a Gd-doped detector, since IBD events can be identified via prompt and delayed coincidences at reactor experiments~\cite{Conrad:2004gw}. In the short-baseline experimental configuration considered in this work, the tagging capabilities are enhanced due to the Gd-doped liquid scintillator (LS) detector technology~\cite{JUNO:2020ijm}. Hence, we assume IBD background rates to be negligible. The primary sources of background in this experimental setup fall into two categories. The first consists of correlated backgrounds induced by cosmic-ray muons, which generate fast neutrons~\cite{JUNO:2024jaw} as well as unstable isotopes including \(^{9}\text{Li}\) and \(^{8}\text{He}\)~\cite{Jollet:2019syr}. The second is an accidental background resulting from random coincidences between a single neutron capture event and signals from natural radioactivity~\cite{Conrad:2004gw}. Regarding cosmic muon related contaminants, we follow a muon veto strategy and consider a 90\% efficiency with a fiducial volume (FV) cut \footnote{As in the case of the muon contamination, accidental backgrounds include the same efficiency and FV cuts as in Ref.~\cite{Li:2022wqc}.} (which corresponds to about 1 tonne of GdLS) as for IBD events~\cite{JUNO:2020ijm}. In this study, an energy cut of 0.2 MeV was applied in both cases.

\begin{figure}[H]
    \centering
    \includegraphics[width=0.75\linewidth]{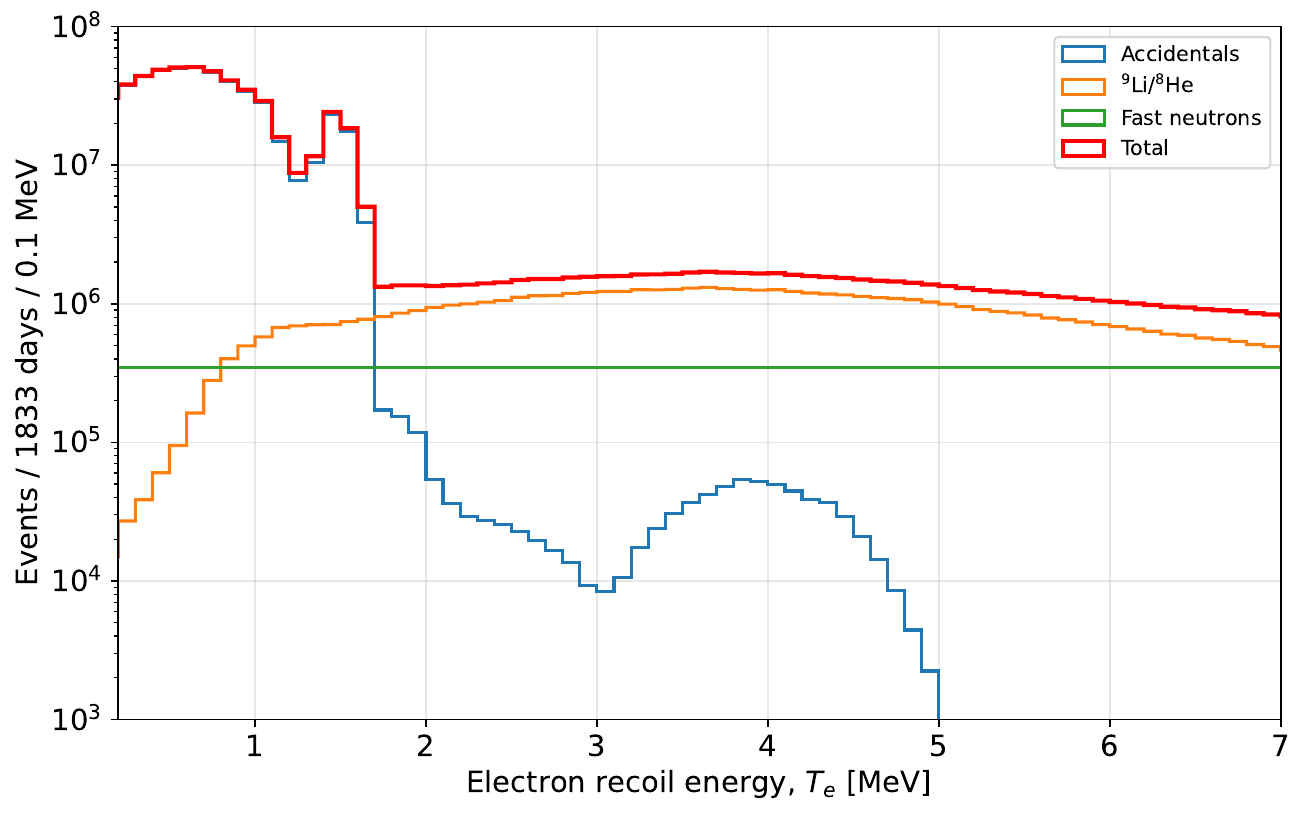}
    \caption{Expected background events in the TAO-like configuration. Accidental backgrounds are shown in blue, \(^{9}\text{Li}\) and \(^{8}\text{He}\) contaminants in orange, and fast neutrons in green. Refer to the text for a detailed explanation.}
    \label{fig:backg}
\end{figure}
In Fig.~\ref{fig:backg} we show the expected background event rates for a TAO-like setup with 25~meters of water equivalent (m.w.e.) shielding~\cite{JUNO:2020ijm, Brdar:2024lud}, assuming a data-taking exposure of 2000 days, with a duty cycle factor of $\tau = 11/12$. The total event rate, displayed in red, represents the sum of all background contributions. Accidental backgrounds (blue), \(^{9}\text{Li}\) and \(^{8}\text{He}\) contaminants (orange), and fast neutrons (green) are each shown as separate components. Here, accidental backgrounds include singles from the \(^{40}\)K isotope and contributions from the \(^{238}\)U and \(^{232}\)Th chains~\cite{Conrad:2004gw}. Fast neutrons were incorporated as in Ref.~\cite{JUNO:2024jaw}, while for \(^{9}\text{Li}\) and \(^{8}\text{He}\) contaminants we follow the methodology presented in Ref.~\cite{Jollet:2019syr}. For all backgrounds, we rescaled the corresponding events according to the expected overburden at the TAO configuration. Besides, for electron recoil energies below $T_e \simeq 2$~MeV, accidental backgrounds constitute the primary contribution, while above this threshold, muon-related contaminants dominate. However, certain techniques could be applied to mitigate muon-induced contaminants at the TAO configuration, following those explored for the JUNO detector~\cite{Nairat:2025sju}.

\subsection{Signal prediction}
\label{subsec:signal}
For the purpose of assessing the predicted (signal) E$\nu$ES events, we consider a baseline (source-to-detector distance) of $L =$ 44 m from a reactor core; the assumed fission fractions of the antineutrino flux $\Phi_\nu (E_\nu)$ are 0.561, 0.076, 0.307, and 0.056 for $^{235}$U, $^{238}$U, $^{239}$Pu, and $^{241}$Pu~\cite{JUNO:2020ijm}, whose predictions rely on the Huber-Mueller model~\cite{Huber:2011wv, Mueller:2011nm}. The corresponding antineutrino flux is given as
\begin{equation}
    \Phi_\nu (E_\nu) = \dfrac{W_{\mathrm{th}}}{ 4\pi L^2 \langle E \rangle} \left( \sum\limits_{\mathrm{iso}} S^{\mathrm{iso}}(E_{\nu}) \right)\;,
\end{equation}
in this analysis, we have considered a mean energy per fission of $\langle E\rangle = 208$ MeV and thermal power $W_{\mathrm{th}} = 4.6$ GW$_{\rm th}$. Unless stated otherwise, we will consider a time exposure of 2000 days of data taking with a duty cycle factor of $\tau = 11/12$, equivalent to 6 calendar years. 

Moreover, in this study, we set up a total of 69 energy bins, uniformly distributed over the energy window of the recoiled electron kinetic energy $T_e$: from $T_e^{\rm min} =$ 0.2 MeV to $T_e^{\rm max} =$ 7 MeV, which corresponds to a bin size of 0.1 MeV. The neutrino energy ranges from $E_{\nu}^{\text{min}} = \frac{1}{2} \big(T_e^{\text{min}} + \sqrt{(T_e^{\text{min}})^2 +2 m_e T_e^{\text{min}}}\big)\simeq0.347$ MeV to $E_{\nu}^{\text{max}} = 10$ MeV.
Therefore, the predicted differential event rates in the absence of detector effects can be computed as (see, e.g., Refs.~\cite{Canas:2016vxp, Brdar:2024lud})
\begin{equation}
    \frac{dN}{dT_e} = N_e~\Delta t~\epsilon \int_{T_e^{\rm min}}^{T_e^{\rm max}} dT_e \int_{E_{\nu}^{\text{min}}}^{E_{\nu}^{\text{max}}} dE_{\nu}\Phi_{\nu}(E_\nu) \frac{d \sigma (T_e, E_\nu)}{dT_e}~,
\end{equation}
here, $N_e$ is the number of electrons in the detector fiducial volume (FV), $E_\nu$ is the neutrino energy, $\Delta t$ is the data taking time including duty factor, $\epsilon$ is the signal efficiency ($90\%$), $d \sigma (T_e, E_\nu) / dT_e$ is the E$\nu$ES differential cross section, \footnote{In this analysis, we are not considering radiative corrections to the E$\nu$ES differential cross section. For an impact of such radiative corrections at short-baseline reactor experiments, see, e.g., Ref.~\cite{Brdar:2024lud}.} $\Phi_{\nu} (E_\nu)$ is the electron antineutrino flux, and $T_e$ is the kinetic energy of the electron. In order to calculate the event rates per energy bin ($i$), we integrate $dN/dT_e$ across a given energy bin. Implementation of detector effects will be described in detail in subsection~\ref{subsec:effects}.

\subsection{Detector response for a liquid scintillator detector}
\label{subsec:effects}
Regarding the detector response, we have considered the impact of LS non-linearity and leakage as well as energy resolution~\cite{Xu:2022mdi}. 
Liquid scintillator (LS) technology, in combination with photo-sensors such as photomultiplier tubes (PMTs), is crucial in neutrino physics, particularly in reactor antineutrino experiments. The energy response of LS detectors is non-linear with respect to the kinetic energies of charged particles, particularly due to ionization quenching as well as Cherenkov radiation~\cite{DayaBay:2019fje}. The main particles involved in the detection of either neutrinos $\nu_{\ell}$ or antineutrinos $\bar{\nu}_{\ell}$ via LS detectors are positions $e^{+}$, electrons $e^-$, and photons ($\gamma$)~\cite{JUNO:2020xtj}. 

Throughout this study, we consider the liquid scintillator nonlinearity (LSNL) of electrons as $T_{\mathrm{vis}}/T_{\mathrm{dep}}$, where $T_{\mathrm{dep}}$ is the deposited energy of electrons in the scintillator, and $T_{\mathrm{vis}}$ is the visible energy defined as the expected reconstructed energy assuming perfect energy resolution.
In this paper, the term LSNL refers to the electron LSNL by default, which is primarily calibrated using gamma calibration sources.
The nonlinearity response of such a detector can be measured with similar calibration sources and procedures as in Daya Bay~\cite{DayaBay:2019fje}, which has a similar scintillator composition.

After implementation of LSNL effects, to obtain the final events in reconstructed visible energy $T_{\rm rec}$, we apply Gaussian resolution smearing $R(T_{\rm vis},~T_{\rm rec})$. Firstly, we have obtained the expected events that deposit energy in the scintillator, $T_{\rm dep}$, which follow a non-linear relation with respect to the visible energy $T_{\rm vis}$, and finally, we apply energy resolution smearing in a sequence of steps given as
\begin{equation}
 T_{\rm dep} \rightarrow  T_{\rm vis} = f_{\rm LSNL}(T_{\rm dep})~T_{\rm dep} \rightarrow R(T_{\rm vis},~T_{\rm rec}) \rightarrow T_{\rm rec}\;.
\end{equation}
Determining the neutrino mass ordering (NMO) requires an outstanding energy resolution of around or greater than 3\% at 1 MeV in the FV. In this study, we parameterize the energy resolution model to electrons within the fiducial volume with a Gaussian energy resolution $R(T_{\rm vis},\,T_{\rm rec})$ as
\begin{equation}
\label{eq:eres}
  R(T_{\rm vis},\,T_{\rm rec})  = \frac{1}{ \sqrt{2 \pi} \sigma_T} \exp{\frac{-(T_{\rm vis} - T_{\rm rec})^2}{2 \sigma^2_T}},\,
\end{equation}
being $\sigma_T =2\%/ \sqrt{T_{\rm vis}\,[\text{MeV}]}$, see, e.g.,~\cite{Xu:2022mdi} and references therein.

For energy leakage, which is the loss of energy due to prompt events depositing part of their energy in non-active volumes, we follow the same strategy as the NMO analysis~\cite{JUNO:2024jaw}, the corresponding energy leakage matrix was continued up to 0.5 MeV below the energy threshold of 0.2 MeV. For this particular neutrino--electron elastic process, leakage effects were found to be sub-leading and much smaller than LSNL and resolution effects.

\section{Systematic uncertainties and statistical analysis}
\label{syst}
\subsection{Systematic uncertainties}
As far as systematic uncertainties considered in this study, we assume a 3\% signal normalization error. Moreover, based on previous assessments from the Daya Bay collaboration~\cite{DayaBay:2017lpf}, we consider a $5\%$ normalization uncertainty on the corresponding muon-related backgrounds, and similarly for the radioactive contaminants. 
Next, we consider LSNL non-linearity systematics for which we follow the standard treatment from Ref.~\cite{JUNO:2024jaw}. This effect is parameterized by the inclusion of one nominal LSNL curve plus four additional pull curves. Besides, To account for the threshold energy uncertainty, in this analysis, we assume a conservative 1\% error on the energy scale of the predicted signal E$\nu$ES events. The energy scale, parameterizes the non-linear scintillator quenching effect of liquid scintillator detectors such as Daya Bay, JUNO and other similar experiments including TAO~\cite{JUNO:2024jaw}. 
Regarding background-related systematics, we have estimated a $\sigma_{r}=1.0\%$ bin-to-bin (b2b) uncorrelated shape uncertainty on radioactive contaminants as well as a $\sigma_{\mu}=1.5\%$ b2b shape uncertainty on muon-related backgrounds, accordingly. These uncertainties are estimated for 6 calendar years of data-taking, during which backgrounds can be precisely characterized using data from an assumed 6-month reactor-off period (see, e.g., Sec~V-C in Ref.~\cite{JUNO:2024jaw}).


\subsection{Data analysis details}
\label{subsec:chisq}
The impact of LSNL is parameterized as~\cite{JUNO:2024jaw}
\begin{equation}
    f^{\mathrm{LSNL}}(T_{\mathrm{dep}}) = f^{\mathrm{LSNL}}_{\mathrm{nominal}}(T_{\mathrm{dep}}) + \sum\limits_{j = 1}^{4} \zeta_j f^{\mathrm{LSNL}}(T_{\mathrm{dep}})_{\mathrm{pull}, j},
\end{equation}
where $\zeta_j = 0 \pm 1$, are the corresponding pull terms of the additional LSNL curves. In order to quantify the statistical significance of our results, we construct a chi-squared function as:
\begin{equation}
\label{eq:chi2}
    \begin{split}
        &\chi^2_{\rm} = \sum^{N_{\rm bin}}_{i=1} \dfrac{(M^{\rm }_i - D^{\rm }_i)^2}{M_i +\sigma_{\rm b2b}^2 B_i^2}  + \Bigg( \frac{1-N^{\rm norm}_{\rm signal}}{\sigma_{\rm signal}} \Bigg)^2 + \Bigg( \frac{1-N^{\rm norm}_{\rm backg}}{\sigma_{\rm backg}} \Bigg)^2 + \Bigg(\frac{1-\eta_{\rm scale}}{\sigma_{\rm scale}}\Bigg)^2 + \chi^2_{\rm penalty}(\text{LSNL})\,, \\
        & M^{\rm }_{i} = \eta_{\rm scale} N^{\rm norm}_{\rm signal} R_{\rm rebin} R_{\rm eres} R_{\rm LSNL} R_{\rm leak}\left(f^{\rm nominal} + \sum\limits_{j =1}^{4}\zeta_j f_{j}^{\rm pull}\right) P_i(\Omega)~+ \\
        &N^{\rm norm}_{\rm backg}R_{\rm rebin} R_{\rm eres} R_{\rm LSNL}\left(f^{\rm nominal}\right) B_i\;.
    \end{split}
\end{equation}
Here, $P_i(\Omega)$ refers to the predicted signal E$\nu$ES events under the parameters $\Omega = (\sin^2 \theta_W$, $\mu_{\nu}$) to be determined, while $B_i$ accounts for the total muon and radioactivity background events accordingly; $D_i = S_i+B_i$ are the total expected events including backgrounds, being $S_i = P_i(\Omega_{\rm true})$. Furthermore, $N^{\rm norm}_{\rm signal}$ and $N^{\rm norm}_{\rm backg}$ are the signal and background normalization nuisance parameters. Besides, $R_{\rm rebin} R_{\rm eres} R_{\rm LSNL} R_{\rm leak}$ is the product of matrices that encodes detector effects: LSNL as well as energy resolution. Moreover, $\sigma_{\rm backg} = 0.05$ and $\sigma_{\rm signal} = 0.03$ account for the corresponding background and signal normalization uncertainties, while $\sigma_{\rm scale} =  0.01$ is the energy scale uncertainty, and $\sigma_{\rm b2b}^2 = \sigma_{r}^2 +\sigma_{\mu}^2$, is the uncorrelated bin-to-bin uncertainty on the background events.
Our projected sensitivities were calculated based on the $\Delta \chi^2 = \chi^2 - \chi ^2_{\text{min}}$ distribution (see, e.g., Refs.~\cite{Blennow:2013oma, JUNO:2024jaw}); we scan over the test parameter $\Omega = \sin^2 \theta_W~\text{or}~\mu_{\nu}$, and map the $\Delta \chi^2$ to the corresponding confidence levels (CL) of the $\chi^2$ distribution considering Wilks' Theorem~\cite{Wilks:1938dza}. In addition, in our $\chi^2$ analysis, we have marginalized the nuisance parameters $\zeta_j, N^{\rm norm}_{\rm signal}, N^{\rm norm}_{\rm backg}$ and $\eta_{\rm scale}$.
\section{Results}
\label{sec:results}
In this section of the manuscript, we present our results for the expected sensitivities to the weak mixing angle, neutrino charge radius, and magnetic moment of neutrinos. The analysis was carried out by two independent groups. Both analysis groups implemented independent software for computing the predicted events at a short-baseline detector, including systematic uncertainties as well as statistical analysis. Both groups carried out detailed cross-checks at all stages of the analysis, from the prediction of the energy spectra to the determination of the sensitivity to the weak mixing angle and magnetic moment of neutrinos. Consistent results were obtained, and only one set of results is presented in this manuscript.  

Common inputs were employed by both groups. These include the antineutrino spectra from the reactors, E$\nu$ES cross-section, detector efficiency, LNSL curves, energy response, expected backgrounds, and other parameters characterizing the performance of the nuclear reactors and antineutrino detectors. In terms of the E$\nu$ES spectra, the relative difference between the groups was at the $\lesssim 2\%$ level in 90\% of the energy range, excluding the high-energy region where event counts are low. A similar agreement was observed for the total predicted signal rates. Compared to the inverse beta decay (IBD) sample, the E$\nu$ES sample suffers from low statistics. Therefore, the main limitation in the expected sensitivities are backgrounds. Our sensitivities in this analysis include the impact of detector effects, such as LNSL and energy resolution; the impact of energy leakage was found to be negligible.  
\begin{figure}[H]
\includegraphics[width=0.65\textwidth]{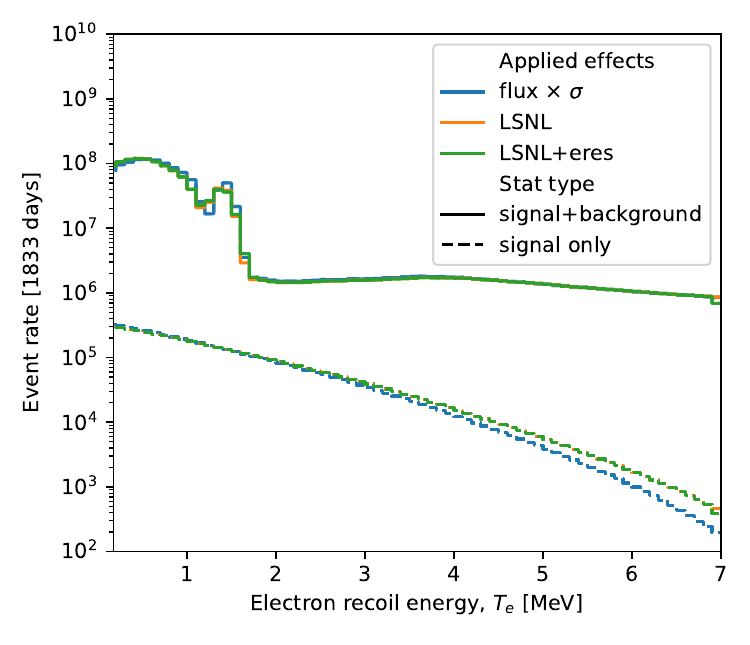}
		\hfill	
 \caption{The expected signal and total event rates comparison with and without detector effects for a short-baseline reactor experimental similar to TAO. Here, dashed lines represent signal-only events while solid lines consider signal plus background events. For instance, the flux times cross section ($\sigma$) rates are shown in blue. Besides, orange lines display the influence of liquid scintillator non-linearity (LSNL). In addition, LSNL plus energy resolutions effects are considered as well (green lines). Refer to the text for a detailed explanation.}
  \label{fig:ratestot}
\end{figure}
In Fig.~\ref{fig:ratestot}, we display the expected signal and background events for a detector at a 44-meter baseline. The dashed lines show the corresponding signal rates with and without detector effects. Moreover, solid lines present the total (signal plus background) rates for the cases with and without detector effects. The flux times cross section ($\sigma$) rates are displayed in blue. Orange lines illustrate the influence of liquid scintillator non-linearity effects. Furthermore, green lines show the combination of LSNL and energy resolution effects.

\subsection{Weak mixing angle sensitivity}
\label{wma}
The weak mixing angle, or Weinberg angle, is an important input within the SM~\cite{Weinberg:1967tq}; it has been accurately measured at high energies by the LEP collider~\cite{ALEPH:2005ab}. However, only a limited number of measurements of this parameter near the MeV scale are available~\cite{Giunti:2014ixa, Erler:2017knj}. For instance, a global assessment of the weak mixing angle from low-energy neutrino experiments can be found elsewhere~\cite{Canas:2016vxp, Barranco:2007ea}. Besides, long-baseline neutrino experiments such as DUNE could determine the weak mixing angle at the percent precision~\cite{deGouvea:2019wav}.

In this subsection of the manuscript, we will assess the weak mixing angle ($\sin^2\theta_W$) sensitivity at a short-baseline reactor experiment. 
\begin{figure}[H]
\includegraphics[width=0.5\textwidth]{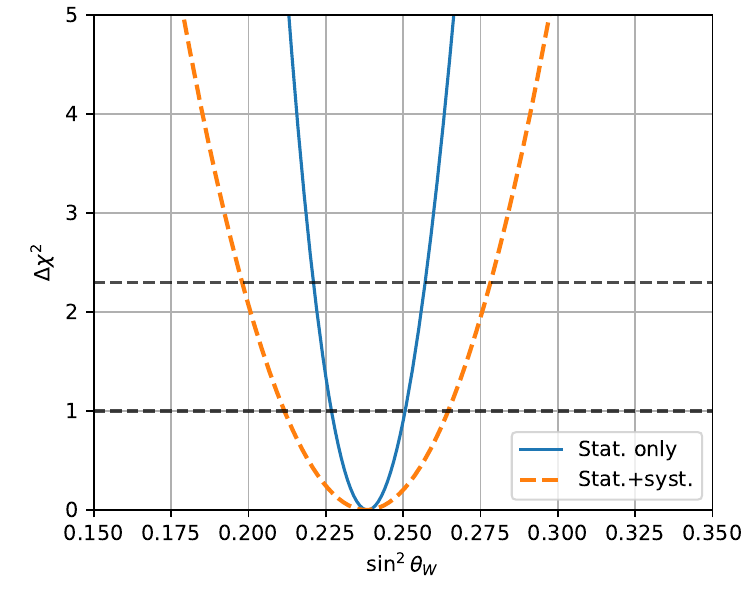}
		\hfill	
 \caption{The expected sensitivity ($\Delta \chi^2$) of the weak mixing angle $\sin^2 \theta_W$ at a 44-meter baseline experiment considering statistics (statistics plus systematics) displayed in blue solid curve (orange dashed curve). The black dashed lines represent the boundaries at $\Delta \chi^2 =2.7$ or 90\% CL ($\Delta \chi^2 =1$ or 1$\sigma$ CL), accordingly. Refer to the text for a detailed explanation.}
  \label{fig:wma}
\end{figure}
In Fig.~\ref{fig:wma}, we display the projected sensitivity to the weak mixing angle $\sin^2 \theta_W$ for a detector at 44 m from the reactor considering statistics (statistics plus systematics) displayed in blue solid curve (orange dashed curve). The black dashed lines represent the boundaries at $\Delta \chi^2 =2.7$ or 90\% CL ($\Delta \chi^2 =1$ or 1$\sigma$ CL), respectively. For instance, the expected 90\% CL sensitivity to the weak mixing angle is:
\begin{equation}
\label{wmasens}
\sin^2 \theta_W =0.239^{+0.037}_{-0.041} \,,
\end{equation}
which is as competitive as other reactor neutrino experiments (see Table~\ref{tab:NWMA} for a comparison). 
\begin{table}[H]
\caption{Current reactor E$\nu$ES and global analysis on $\sin^2 \theta_W$ as well as the projected sensitivity at a short-baseline experiment (shown in parenthesis).}
\begin{center}
\begin{tabular}{llcrc}
\hline \hline
Process &~~Experiment & Measurement (sensitivity) & CL & Ref.\\
\hline \hline
\multirow{4}{*}{E$\nu$ES}
&~~Krasnoyarsk	&$\sin^2 \theta_W = 0.22^{+0.7}_{-0.8} $	&~~~90\%	&\cite{Vidyakin:1992nf}\\
&~~TEXONO	&$\sin^2 \theta_W =0.251 \pm 0.055 $	&~~~90\%	&\cite{TEXONO:2009knm}\\
&~~Global   &$\sin^2 \theta_W = 0.259 \pm 0.025$  &~~~90\%  &\cite{Barranco:2007ea}\\
&~~SBL--1 tonne (44 m)	&($\sin^2 \theta_W =0.239^{+0.037}_{-0.041}$) &~~~90\%	&~~This work\\
\hline \hline
\end{tabular}
\end{center}
\label{tab:NWMA}
\end{table}

\section{Neutrino charge radius}
\label{ncr}
In the SM, neutrinos can have a non-zero charge radius (CR) despite the fact that they are neutral particles~\cite{Bernabeu:2000hf, Bernabeu:2002pd}. 
For instance, the SM prediction of the neutrino CR reads~\cite{Bernabeu:2000hf, Giunti:2014ixa}
\begin{equation}
{\left\langle r_{\nu_{\ell}}^2\right\rangle}_{\mathrm{SM}}=\frac{G_F}{4 \sqrt{2} \pi^2}\left[3-2 \ln \left(\frac{m_{\ell}^2}{M_W^2}\right)\right],
\label{crsm} 
\end{equation}
being $G_F$ the Fermi constant, and $M_W$ and $m_{\ell}$ are the $W$ boson and charged lepton masses, respectively. The numerical evaluation of Eq.~(\ref{crsm}) for each lepton flavor gives~\cite{Bernabeu:2000hf, Bernabeu:2002pd}
\begin{equation}
{\left\langle r_{\nu_{\ell}}^2\right\rangle}_{\mathrm{SM}} \simeq
\begin{cases}
4.1 \times 10^{-33} \mathrm{~cm}^2, & \mbox{ if $ \ell = e $}\,,\\
2.4 \times 10^{-33} \mathrm{~cm}^2, & \mbox{ if $ \ell = \mu $}\,,\\
1.5 \times 10^{-33} \mathrm{~cm}^2, & \mbox{ if $ \ell = \tau $}\,.
\end{cases}
\end{equation}
Different experimental limits to the neutrino change radius have been reported in the literature (see, e.g.,~\cite{Giunti:2024gec} and references therein); for the electron and muon flavors, these constraints are only about an order of magnitude distinct from the SM prediction and could be the first electromagnetic properties of neutrinos to be measured by upcoming experiments.

In order to set limits on the neutrino charge radius (CR) at a short-baseline experiment, we follow the standard approach for the neutrino charge radius (see, e.g., Refs.~\cite{Giunti:2015gga, Brdar:2024lud, AtzoriCorona:2024rtv, Cadeddu:2018dux}), where the CR contribution is included through a shift of $\sin^2 \theta_W$ \footnote{In this analysis, we consider only the charge radius contribution added to the tree-level elastic neutrino-electron scattering (E$\nu$ES) cross section.},
\begin{equation}
\label{eq:wmashift}
    \sin^2 \theta_W \rightarrow \sin^2 \theta_W \left(1 + \frac{1}{3} M_W^2 \langle r_{\nu_{\ell}}^2 \rangle \right),
\end{equation}
at the level of the differential E$\nu$ES cross section (Eq.~\ref{xsc}). A similar strategy was employed in the past by the TEXONO collaboration~\cite{TEXONO:2009knm}. However, as demonstrated by the authors of Ref.~\cite{Brdar:2024lud}, to fully assess the sensitivity and discovery potential of the neutrino charge radius at upcoming reactor neutrino experiments, not only a complete set of one-loop level radiative corrections but also the inclusion of additional detection channels will be necessary.

Furthermore, for ultra-relativistic neutrinos at low-momentum transfer ($q^2$), the effective neutrino charge radius $\langle r_{\nu_{e}}^2 \rangle_{\text{eff}}$ could include contributions from the anapole moment~\cite{Giunti:2024gec}. Therefore, in deriving our bounds, the limits on the neutrino charge radius $\langle r_{\nu_e}^2 \rangle$ can be interpreted as limits on the effective neutrino charge radius $\langle r_{\nu_{e}}^2 \rangle_{\text{eff}}$. A summary of bounds on the neutrino CR from several experimental measurements, as well as the expected limits at a 44-meter baseline experiment, is shown in Table~\ref{tab:NCR}.

\begin{table}[H]
\caption{Reactor and accelerator E$\nu$ES limits on the neutrino charge radius $\langle r_{\nu_{e}}^2 \rangle$ and expected limit at a short-baseline experiment.}
\begin{center}
\begin{tabular}{llcrc}
\hline \hline
E$\nu$ES & Experiment & Limit $[10^{-32}\,\text{cm}^2]$ & CL & Ref.\\
\hline \hline
\multirow{3}{*}{Reactor $\bar\nu_e$}
&Krasnoyarsk	&$|\langle r_{\nu_{e}}^2 \rangle|<7.3$		&90\%	&
\cite{Vidyakin:1992nf}\\
&TEXONO		&$\langle r_{\nu_{e}}^2 \rangle\in(-4.2, 6.6)$	&90\%	&
\cite{TEXONO:2009knm}\\
&SBL--1 tonne (44 m)	&$\langle r_{\nu_{e}}^2 \rangle\in(-5.52, 6.35)$	&90\%	&
This work\\
&SBL--1 tonne (44 m)	&$\langle r_{\nu_{e}}^2 \rangle^{*}\in(-5.40,6.22)$	&90\%	&
This work\\
\hline
\multirow{2}{*}{Accelerator $\nu_e$}
&LAMPF		&$\langle r_{\nu_{e}}^2 \rangle\in(-7.12,10.88)$	&90\% &
\cite{Allen:1992qe}\\
&LSND		&$\langle r_{\nu_{e}}^2 \rangle\in(-5.94,8.28)$	&90\%	&
\cite{LSND:2001akn}
\\
\hline \hline
\end{tabular}
\end{center}
\label{tab:NCR}
\end{table}

In setting the limits for the neutrino charge radius, $\langle r_{\nu_{\ell}}^2 \rangle$, we consider the Standard Model (SM) best-fit value of the weak mixing angle at low momentum transfer ($Q$), namely $\sin^2\theta_W^{\rm SM} = 0.2385$~\cite{Tiesinga:2021myr}. We employ two approaches: the first approach directly maps the limits obtained from the weak mixing angle sensitivity to the charge radius (CR), as previously discussed. We find $\langle r_{\nu_{e}}^2 \rangle \in (-5.52, 6.35)\times10^{-32}\rm cm^2$ at 90\% confidence level (CL). 
In the second approach, we subtract the uncertainty from electroweak (EW) measurements~\cite{AtzoriCorona:2024vhj} to isolate solely the uncertainty from neutrino-related processes, obtaining: $\langle r_{\nu_{e}}^2 \rangle^{*} \in (-5.40, 6.22)\times10^{-32}\rm cm^2$ at 90\% CL. Both limits are included in Table~\ref{tab:NCR} accordingly.

\subsection{Effective neutrino magnetic moment sensitivity}
\label{nmm}
In the SM, neutrinos are massless, and they cannot have a magnetic or electric dipole moment. However, massive (Dirac or Majorana) neutrinos can develop a magnetic moment at the one-loop level~\cite{Fujikawa:1980yx}. In the simplest extension of the SM with three massive Dirac neutrinos, the diagonal magnetic moments are strongly suppressed by the smallness of neutrino masses~\cite{Giunti:2024gec}
\begin{equation}
    \mu_{\nu}^{\text{D}} = \frac{3e G_F m_i}{8 \sqrt{2}\pi^2} \simeq 3\times10^{-20} \big(\frac{m_i}{0.1~\text{eV}} \big) \mu_{\text{B}}~,
\end{equation}
here, $e$ is the elementary charge, $G_F$ the Fermi constant, $m_i$ the neutrino mass and $\mu_{\text{B}} \equiv e /(2 m_e)$ the Bohr magneton. 
Experimental searches for neutrino magnetic moments have been extensively investigated via the E$\nu$ES process. Regarding reactor neutrino experiments, the most stringent limit on $\mu_{\nu_e}$ has been obtained by the GEMMA collaboration~\cite{Beda:2012zz}. For instance, the neutrino magnetic moment ($\mu_{\nu_\ell}$) contributes to the neutrino--electron scattering cross section as~\cite{Vogel:1989iv}
\begin{equation}
\label{mmxsec}
     \frac{d \sigma (\mu_{\nu_\ell})}{dT_e}  = \frac{\pi \alpha^2 }{m_e^2} \abs{\frac{\mu_{\nu_\ell}^2}{\mu_{\rm B}}} \Big( \frac{1}{T_e} -\frac{1}{E_\nu}\Big)~,
\end{equation}
being $\alpha$ the fine-structure constant, $T_e$ the kinetic energy of the recoiled electron, and $E_\nu$ the corresponding neutrino energy. Due to its $1/T_e$ functional dependence in the cross section, the sensitivity to the magnetic moment is enhanced for low values of $T_e$ at given $E_\nu$.
\begin{figure}[H]
\includegraphics[width=0.54\textwidth]{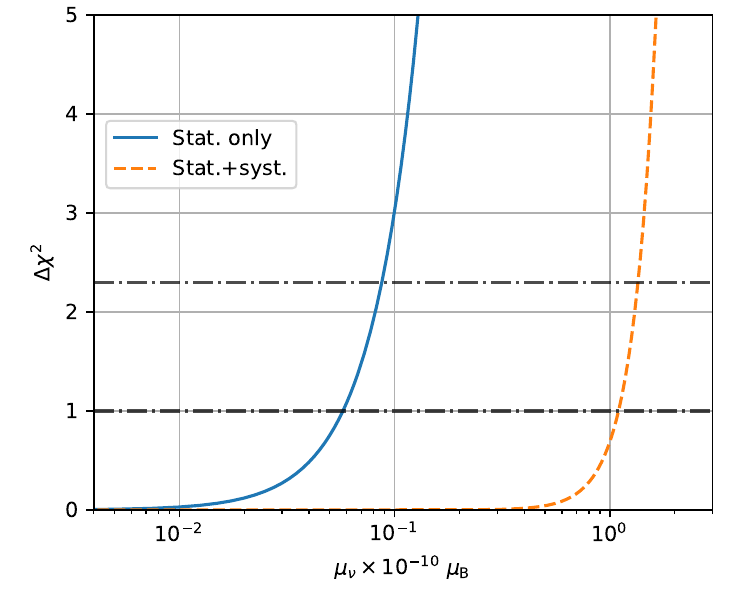}
		\hfill	
 \caption{The expected sensitivity ($\Delta \chi^2$) to the neutrino magnetic moment $\mu_{\nu_{e}}$ for a short-baseline reactor experiment considering statistics (statistics plus systematics) displayed in blue solid curve (orange dashed curve). The black dashed lines represent the boundaries at $\Delta \chi^2 =2.7$ or 90\% CL ($\Delta \chi^2 =1$ or 1$\sigma$ CL), accordingly. Refer to the text for a detailed explanation.}
  \label{fig:mmtao}
\end{figure}
In Fig.~\ref{fig:mmtao}, we display the projected sensitivity to the neutrino magnetic moment $\mu_{\nu_{e}}$ for a detector at 44 m from the reactor considering statistics (statistics plus systematics) displayed in blue solid curve (orange dashed curve). The black dashed lines represent the boundaries at $\Delta \chi^2 =2.7$ or 90\% CL ($\Delta \chi^2 =1$ or 1$\sigma$ CL), respectively. Hence, the expected 90\% CL sensitivity to the neutrino magnetic moment ($\mu_{\nu_e}$) considering all systematic effects is
\begin{equation}
\label{wmasens}
\mu_{\nu} < 1.2 \times 10^{-10} \mu_{\rm B} \,.
\end{equation}
Compared with bounds from other reactor neutrino experiments, a short-baseline setup may be able to set competitive limits. Nevertheless, it is important to note that in order to obtain a more robust estimate of the effective neutrino magnetic moment bound, further studies on non-linearity effects at low energies and improved background reduction would enhance the aforementioned sensitivities. Regarding low-energy non-linearity effects, an alternative approach could be to deploy sub-MeV calibration sources for such detectors~\cite{Takenaka:2024ctk}.

However, reactor experiments are not as competitive as dark matter direct detection experiments, for which the sensitivity reach on the neutrino magnetic moment is $\mu_{\nu_\ell} \lesssim 1.3 \times 10^{-11} \mu_\text{B}$~\cite{Giunti:2024gec}. In Table~\ref{tab:NMM}, we show the projected sensitivity to the neutrino magnetic moment at a short-baseline experiment, as well as present bounds from other neutrino experiments.

\begin{table}[H]
\caption{Bounds on the effective magnetic moment of neutrino from several neutrino experiments. The sensitivity for a short-baseline setup is shown in parenthesis.}
\begin{center}
\begin{tabular}{llcrc}
\hline \hline
E$\nu$ES & Experiment & Limit (sensitivity) $[\times 10^{-11} \mu_{\text{B}}]$ & CL & Ref.\\
\hline \hline
\multirow{8}{*}{Reactor $\bar\nu_e$}
&Krasnoyarsk	&$\mu_{\nu_e} < 24 $	&~~90\%	&\cite{Vidyakin:1992nf}\\
&Rovno			&$\mu_{\nu_e} < 19 $	&~~95\%	&\cite{Derbin:1993wy}\\
&MUNU			&$\mu_{\nu_e} < 9  $	&~~90\%	&\cite{MUNU:2005xnz}\\
&TEXONO			&$\mu_{\nu_e} < 7.4 $	&~~90\%	&\cite{TEXONO:2006xds}\\
&GEMMA			&$\mu_{\nu_e} < 2.9 $	&~~90\%	&\cite{Beda:2012zz}\\
&CONUS          &$\mu_{\nu_e} < 7.5 $    &~~90\%  &\cite{CONUS:2022qbb}\\
&SBL--1 tonne (44 m) &($\mu_{\nu_e} < 12 $)	&~~90\%	&~~This work\\
CE$\nu$NS + E$\nu$ES
& Dresden-II~\cite{Colaresi:2022obx}
& $\mu_{\nu_e} < 21 $
& 90\%
& \cite{Coloma:2022avw, AtzoriCorona:2022qrf}
\\
\hline \hline
\end{tabular}
\end{center}
\label{tab:NMM}
\end{table}

\begin{table}[H]
\caption{Short-baseline reactor neutrino experimental configurations.}
\begin{center}
\begin{tabular}{lccccc}
\hline \hline
Experiment & Baseline [m] & Thermal Power [GW$_{\text{th}}$] & $\overline{\nu}_e$ flux [$\overline{\nu}_e$/(cm$^2\cdot$s)] & Fiducial Volume \\
\hline \hline
TEXONO       & 28   & 2.9  & $6.4 \cdot 10^{12}$ & 187 kg \cite{TEXONO:2006xds} \\
SBL--1 tonne & 44   & 4.6  & $4.1 \cdot 10^{12}$ & 1-tonne (this work)  \\
Krasnoyarsk  & 20.5 & 1.7 (estimated) & $7.0 \cdot 10^{12}$ & 103 kg \cite{Vidyakin:1992nf}\\
Rovno        & 15    & 3.0 (estimated) & $2.0 \cdot 10^{13}$ & 75 kg \cite{Derbin:1993wy} \\
MUNU         & 18   & 2.75 & $2.0 \cdot 10^{13}$ & 10 m\textsuperscript{3} \cite{MUNU:2005xnz} & \\
GEMMA        & 13.9 & 3.0  & $2.7 \cdot 10^{13}$ & 1.5 kg \cite{Beda:2012zz}\\
CONUS        & 17.1 & 3.9  & $2.3 \cdot 10^{13}$ & 1.0 kg \cite{CONUS:2022qbb}\\
Dresden II   & 10.4 & 2.96 & $4.8 \cdot 10^{13}$ & 3 kg \cite{Colaresi:2022obx, Coloma:2022avw, AtzoriCorona:2022qrf}\\
\hline \hline
\end{tabular}
\end{center}
\label{tab:reactor_config}
\end{table}
Finally, in Table~\ref{tab:reactor_config}, we summarize the key characteristics of the short-baseline reactor neutrino experiments discussed in this work, including their baselines, thermal power, antineutrino fluxes, and target masses.
\section{Conclusions}
\label{sec:conclusion}
In this paper, we have studied the potential to determine the weak mixing angle and neutrino charge radius at a short-baseline reactor experiment through $\bar{\nu}_e-e^{-}$ elastic scattering with neutrinos from a nuclear reactor. First, we obtain the projected sensitivity to the weak mixing angle, $\sin^2 \theta_W$. The expected sensitivity for a detector at 44-meter baseline including all detector effects and systematic uncertainties (Fig.~\ref{fig:wma}), was estimated to be $\sin^2 \theta_W =0.239^{+0.037}_{-0.041}$ at 90\% CL ($\Delta \chi^2 =2.7$). Furthermore, based on the sensitivity to $\sin^2 \theta_W$, we set the expected bounds on the neutrino charge radius $\langle r_{\nu_e}^2 \rangle$ for such an experimental setup (see Table~\ref{tab:NCR}).

Besides, in this analysis, we assessed the sensitivity to the neutrino magnetic moment $\mu_{\nu_e}$ (Fig.~\ref{fig:mmtao}); the expected 90\% CL limit was found to be $\mu_{\nu_e} < 1.2 \times 10^{-10}~\mu_{\text{B}}$. This bound is comparable with previous bounds from other reactor neutrino experiments (see Table~\ref{tab:NMM}). 

In both cases, it was found that the sensitivity is limited by backgrounds, and LSNL systematics. As far as energy leakage and scale uncertainty, their impact on the sensitivities were found to be negligible. Regarding the magnetic moment limit, in order to properly assess the impact of non-linearity effects near the energy threshold, further studies of calibration sources at the sub-MeV region will be beneficial.

Improvements on the final-state direction reconstruction of the electrons, as well as particle identification of the cosmogenic and radioactive backgrounds, can not only enhance the sensitivity to the weak mixing angle as well as the neutrino magnetic moment at short-baseline reactor experiments.


\section*{Acknowledgments}
We would like to thank Cecile Jollet for kindly providing the publicly available simulation ROOT files corresponding to the decays of cosmogenic nuclei backgrounds, $^{9}$Li and $^{8}$He. We gratefully acknowledge support from the National Key R\&D Program of China under grant No.~2024YFE0110500. Maxim Gonchar and Vitalii Zavadskyi are supported in the framework of the State project ``Science'' by the Ministry of Science and Higher Education of the Russian Federation under the contract 075-15-2024-541. We acknowledge the anonymous referee for the comments and suggestions that have helped us to improve our manuscript.

This paper represents the views of the authors and should not be considered a JUNO collaboration paper.

\appendix

\section{Differences between the analyses}
In this section, we present the main differences among two independently performed analyses, namely group A and B.
The prediction of the expected spectrum follows the methods described in Section~\ref{sec:method}. 
The process of prediction proceeds from neutrino energy $E_\nu$ to reconstructed electron energy $T_{\rm rec}$ follows from~\cite{JUNO:2024jaw}
\begin{equation}
    E_\nu \rightarrow T_e(E_\nu,~\text{Kinematics}) \rightarrow T_{\rm dep} (T_e, ~\text{Leak}) \rightarrow T_{\rm vis}(T_{\rm dep}, \text{LSNL}) \rightarrow T_{\rm rec} (T_{\rm vis}, \text{Resolution}). 
\end{equation}
The first group A, considered analytical implementation of leakage, and LSNL effects within the energy resolution function (see, e.g., Ref.~\cite{Li:2013zyd}). The second group B, implemented LSNL and leak effects via matrix formalism, both effects were applied separately from the energy resolution matrix. Relative difference among groups was found to be less than 1\%. As far as statistical methods, both groups employed a Pearson's definition of the $\Delta \chi^2$ function~\cite{JUNO:2024jaw}. Overall, results among groups were found to be consistent.

\end{document}